# A new type of mirror symmetry in the set of protein amino acids

Miloje M. Rakočević

University of Niš, Faculty of Sciences and Mathematics, Department of Chemistry,
Višegradska 33, 18000 Niš, Serbia

**Abstract.** In several previous works, I presented the mirror symmetry in the set of protein amino acids, expressed through the number of atoms. Here, however, the same thing is shown but over the number of nucleons and molecules mass. Compared to the previous version of the paper, minimal changes have been made, and Display 2 as well as Figures 3 and 4 have been added.

| Canonical AAs mirroring ||||
|---|---|---|---|
| Atom number || Nucleon numb. ||
| 02 | 20 | 981 | 189 |
|  | 11 |  | 100 |
| 13 | 31 | 982 | 289 |
|  | 11 |  | 010 |
| 24 | 42 | **992** | **299** |
|  |  |  | 010 |
| 16 | 61 | 982 | 289 |
|  | 11 |  | 100 |
| 05 | 50 | 981 | 189 |

Synopsis of **Graphical abstract**: Records of the number of atoms and nucleons as *original* and the *image* in the mirror. Reducing the distances between the quantities of the number of atoms and/or of nucleons to the level "filled" exclusively with units, directly supports Shcherbak's hypothesis on the "analogies with quantum physics" (Shcherbak, 1994).

The new type of the mirror symmetry, in the set of protein amino acids (AAs), which is the matter in this brief communication, was go in the form of a hint in a previous my work (MMR, 2004, Table 1, p. 223).[1] But then it was not allowed to say something like that [mirror symmetry of the numbers ?!], because it would be declared as numerology.[2] Now when that symmetry is the factual state, expressed though the number of atoms (Box 1), it makes sense to represent it also through the number of nucleons (Table 1 in relation to Figure 1), and through molecules mass (Tables 2, 3 and 4).[3]

We look now at Tables 1 and 2 together with Figure 1 and Survey 1. If we place a mirror at the middle row of Table 1, either from above or from below, we always see one and the same sequence-image; for the number of nucleons: 299, 289, 189 and for the molecule mass: 596, 586, 485, as indicated in Figure 1.[4] This is directly obvious mirror

---

[1] The word is about a new type of mirror symmetry, in relation to enantiomer symmetry of amino acid molecules (MMR, 2018a, Solution 1, p. 42).

[2] It is the fact, however, that this new type of symmetry I have presented in some of the previous papers published only in preprint form; also once in the official journal, in a regular article (MMR, 2018a, at the end of page 38, in relation to Figure 4, p. 37.).

[3] We are talking here about the set of standard protein amino acids that have the status of "canonical amino acids" because their incorporation into proteins is mediated by the genetic code.

[4] In the middle row, in Table 1 and Figure 1, respectively, the number of nucleons 299 and the quantity of mass 596 irresistibly "invoke" the third perfect number: the quantity 596 because of the difference 596 -

symmetry, but it is existing a hidden one. The full meaning of balancing and nuancing, expressed through distinctions and differences between quantities, it appears only when the original and/or the mirror image of the double value of the third perfect number is included: the image as in Table 2, and original as in Table 3; both Tables taken together as a single system).[5]

**Remark 1.** One can note in Table 2, in the area of the quantities of molecules mass, the difference **100** (bold) in the middle row, and in area of nucleons number the difference **000** (bold). The difference of difference is 100. On the other hand, in Survey 1 we find that the difference between the sums and the sums in Table 2 is also 100. As we see the principle of self-similarity is at work also here as well.

**Box 1.** *Mirror symmetry through the number of atoms as a unity of numerical quantities and chemical qualities*

| | |
|---|---|
| Start: a unique sequence of Periodic System of Numbers (PSN) in front of a mirror, with superposition; then "crossing" with a unique 6-bit binary tree sequence (Box 2): 010/101 as 2 / 5 (3 x 2 and 2 x 5): <br><br> 00-11-22 / 22-11-00 → **00-11-22-11-00** → 02, 13, 24, 16, 05 <br><br> Then mirroring with the result as in Display 1 and Fig. 2 (MMR, 2019, Fig. 2, p. 17 and Tab. A3, p. 30; 2021, Fig. 2, p. 55 and Fig. A2, p. 88). PSN as in (MMR, 2019, Fig. A1, p. 28). | **Display 1**. The unity of chemism and semiosis (I) <br><br> (G 01 + A 04) + (**N 08** + **D 07**) = 20 <br> (V 10 + P 08) + (**S 05** + **T 08**) = 31 <br> (I 13 + L 13) + (**C 05** + **M 11**) = 42 <br> (K 15 + R 17) + (**F 14** + **Y 15**) = 61 <br> (Q 11+ E 10) + (**W 18** + **H 11**) = 50 |
| According to our idea, Display 1 should be proof of our prediction (given at the end of the paper) that the Genetic code represents the unity of chemism and semiosis. The number of atoms in each of the five rows of AAs represents a mirror image of the number that arises in a specific crossing of PSN and the 6-bit binary tree. And the arrangement according to chemism is as follows. In the left (un-bolded) decade, AAs are sorted into pairs by chemical similarity: 1. "G" has the simplest possible side chain (it has only one hydrogen atom), and the side chain of "A" has one carbon atom in addition to three hydrogen; 2. "V" has the simplest semi-cyclic hydrocarbon chain and "P" has the simplest cyclic one; 3. "I" and "L" are the only two AAs (in this set) that are structural isomers; 4. "K" and "R" are the only two AAs with amino group in the side chain; 5. the fifth pair of the first decade forms a chemical unity (by similarity) with the 1st pair of the second decade: the only two AAs with the carboxyl group in the side chain ("E" and "D") with their amides, "Q" and "N", respectively; 2nd pair of second decade consists of only two AAs with the hydroxyl group in the side chain; 3rd pair are the only two sulfur AAs; 4th pair are the only two carbocyclic aromatic AAs; and 5th pair are the only two heterocyclic aromatic AAs. ||

---

496 is equal to 100, and the number 299 because that record is actually a mirror image of the double value of the third perfect number.

[5] That perfect and friendly numbers are determinants of the genetic code I have repeatedly explained, especially in a monograph from 25 years ago (MMR, 1997b, pp. 60-64). www.rakocevcode.rs

**Display 2**. The unity of chemism and semiosis (II)

| 01G 01 | 11N 08 | 06L 13 | 16M 11 | | 01G 01 | 02A 04 | 11N 08 | 12D 07 |
|---|---|---|---|---|---|---|---|---|
| 02A 04 | 12D 07 | 07K 15 | 17F 14 | | 03V 10 | 04P 08 | 13S 05 | 14T 08 |
| 03V 10 | 13S 05 | 08R 17 | 18Y 15 | | 05I 13 | 06L 13 | 15C 05 | 16M 11 |
| 04P 08 | 14T 08 | 09Q 11 | 19W 18 | | 07K 15 | 08R 17 | 17F 14 | 18Y 15 |
| 05I 13 | 15C 05 | 10E 10 | 20H 11 | | 09Q 11 | 10E 10 | 19W 18 | 20H 11 |

On the right side the arrangement of AAs is the same as in Display 1 (20 AAs as two decades), except that the order of amino acids (in the similarity), from the first to the last, is indicated here. (Cf. MMR, 2019, Table 2, p. 14.) The AAs decades placed on the left are divided into quintets: the first with the third, and the second with the fourth. Such an arrangement shows the unity of chemism and semiosis in a different way (different from this presented in Box 1), as it is represented in Figure 3. Furthermore, Figure 4 shows a more complete unity of chemism and semiosis through the chemical properties of molecules, the number of atoms in them, as well as through ordinal number of AAs in the system of chemical similarity of molecules as such.

**Box 2.** *The determination of the series of the numbers 0-63 on the 6-bit binary tree*

```
        /00 - 07/08 - 15/16 - 23/24 - 31//32 - 39/40 - 47/48 - 55/56 - 63/
           28      92     156    220    284     348    412     476
              64      64      64     64     64      64      64

        /00 - 07/00 - 15/00 - 23/00 - 31//00 - 39/00 - 47/00 - 55/00 - 63/
           28     120     276    496    780    1128   1540    2016
              92     156     220    284    348     412     476
```

"The determination of the series of the numbers 0-63 [on the 6-bit binary tree]. When we look closely into the structure of the sequence 0-63 of the series of the natural numbers we come to the obvious and self-evident explanation of the reason why the genetic code must be six-bit code, no matter if it is the manifestation in the form of the Gray Code model (Swanson, 1984, p 188), or it is in the form of the Binary tree (Rakočević, 1994, p. 38) [Rakočević, 1998, Fig. 1, p. 284]. There must be 8 codon i.e. amino acid classes. The structure of the sequence 0-63 is strictly determined by third perfect number (496) and the sum cbonsisted of the first pair of the friendly numbers (220 + 284 = 504) …" (Rakočević, 1997b, Fig. 7, p. 60) www.rakocevcode.rs

**Remark 2.** About "strange result" in Figure 1, where from differences of the quantities of molecules mass and the number of nucleons follows the number of atoms in the AAs molecules it is difficult to give any explanation. Instead that, we give only the facts. In the previous paper (MMR, Table 3a, p. 224) we showed that in the first two columns of GCT there are 196 + 100 = 296 atoms; the other two columns have 154 + 144 = 298 atoms. It is also indicated that in the first two columns there are 297–1 and in the other two columns there are 297 + 1 atoms.

**Remark 3.** The three equations we find reviewed in Survey 2 (very bottom, right) are also found in the standard GCT (MMR, 2004, Table 3a, p. 224).

\*

Our expectations are that over time it will turn out that the presented results are strong support for two hypotheses and two predictions. The first hypothesis is the hypothesis of Vladimir Shcherbak, according that "The laws of additive-positions of numbers … have analogies with quantum physics" (Shcherbak, 1994, p. 476, last passage). The second hypothesis is our hypothesis given in a previous paper, whose this brief communication is a specific supplement: "Hypothesis on a complete [prebiotic] genetic code" (MMR, 2004, Section 7.1, p. 231).

The first prediction we give here for the first time boils down to the expectation that future researches will show that the view of mirror symmetry, presented here, has a direct or indirect connection with the choice of only the left amino acid enantiomers in the act of the generating of life.

According to the second prediction, it is expected that future researches will show that each of the mirror images of quantities (of number of atoms and / or nucleons, as well as of the sums of molecules mass of amino acids), presented here, can be considered as "signifier" of the "signified" in a semiotic sense, in the manner as we presented in the previous paper for Rumer's system of nucleotide doublets.[6] In such a case, "signifier" represents a mirror image, and "signified" is the corresponding chemical entity.

---

[6] MMR, 2018a, pp. 31-31: "Rumer (1966) suggests that encoding by dinucleotide aggregations is mediated by 'grammatical' formalism (the relation between words and the root of the word), semantics (one-meaning and multy-meaning codon families) and by semiology, i.e. semiotics (the classification of nucleotide doublets after the number of their hydrogen bonds which appear here as 'signifiant' and 'signifié' (signifier and signified) at the same time, that is as their unity (De Saussure, 1985, p. 99))."

# Tables

**Table 1.** *"The harmonic structure with two 'acidic' and three 'basic' amino acid quartets"*

| Amino acids | | | | n | Molecule mass |
|---|---|---|---|---|---|
| D 59 | N 58 | A 15 | L 57 | 189 | 485.49 ≈ 485 /486 |
| R 100 | F 91 | P 41 | I 57 | 289 | 585.70 ≈ 586 /586 |
| K 72 | Y 107 | T 45 | M 75 | **299** | 595.71 ≈ **596** /596 |
| H 81 | W 130 | S 31 | C 47 | 289 | 585.64 ≈ 586 /586 |
| E 73 | Q 72 | G 01 | V 43 | 189 | 485.50 ≈ 485 /486 |
| | | | | 1255 | 2738.04<br>2738 = 2 x (037 x 037) |

Minimally reduced and technically adjusted Table from previous work (MMR, 2004, Table 1, p. 223). Designation (n): The number of nucleons within 20 AAs side chains, calculated from the first, the lightest nuclide (H-1, C-12, N-14, O-16, S-32); One can notice that molecule mass within five rows is realized through the same logic-patterns of notations as the first nuclide, i.e. isotope. [*Note*: Integer upper values of molecules mass in all five rows end with the number 6, and the lower with the number 5 (485 + 585, etc.). The sum of the five lower values (485 + 585 +595 + 585 + 485) equals 2735, which number is the same as the number of nucleons in the whole molecules: in their "bodies", i.e. side chains (1255) and in "heads", i.e. amino acid functional groups (1480).]

**Table 2.** *The number of nucleons and the quantities of molecules mass in Table 1, in relation to the third perfect number and the mirror image of its double value*

| 992 - 0<u>1</u>1 | 981 / 189 |  |  | (496 - 0<u>1</u>0) | 48<span style="color:red">6</span> | 485 |  |
|---|---|---|---|---|---|---|---|
|  |  | *<u>1</u>00* |  |  |  |  | *<u>1</u>01* |
| 992 - 0<u>1</u>0 | 982 / 289 |  |  | (496 + 0<u>9</u>0) | 58<span style="color:red">6</span> | 586 |  |
|  |  | *0<u>1</u>0* |  |  |  |  | *0<u>1</u>0* |
| 992 ± **000** | 992 / 299 |  |  | 496 + **<u>1</u>00** | 59<span style="color:red">6</span> | 596 |  |
|  |  | *0<u>1</u>0* |  |  |  |  | *0<u>1</u>0* |
| 992 - 0<u>1</u>0 | 982 / 289 |  |  | (496 + 0<u>9</u>0) | 58<span style="color:red">6</span> | 586 |  |
|  |  | *<u>1</u>00* |  |  |  |  | *<u>1</u>01* |
| 992 - 0<u>1</u>1 | 981 / 189 |  |  | (496 - 0<u>1</u>0) | 48<span style="color:red">6</span> | 485 |  |
| 042 |  | *220* |  | 300 |  |  | *222* |

The sums of differences (the sums of absolute values) correspond to two numbers (220 and 300) in the "logical square" in Figure 2 [(0). 204, **(1). 220**, (2). 284, **(3). 300**]. The other two vertices off the logic square, the numbers 284 and 204, exist through the relation with the golden mean, as shown in Survey 1, below right, in relation to Survey 2. If total molecules mass (M) follows according to column "485", then M = **2** x (037 x 037) (**2738**); if according to column "486", then M = 1 x (1370 + 1370) (**2740**) (Cf. shaded area in Table 4). If both results are taken into account, then the "hidden" mirror symmetry of Dirac's type <u>1</u>37 / <u>0</u>37 (0 vs 1 as electron vs positron) is also revealed. The degree of freedom between the integer in these two columns and the original non-integer, given in Table 1, was determined by the isotope contribution of the chemical elements.

**Table 3.** *The number of nucleons and the quantities of molecules mass in Table 1, in relation to the third perfect number (for nucleons) and its double value (for molecules mass)*

| | | | | | |
|---|---|---|---|---|---|
| 496 - 30**7** | 189 | | (992 - 507) | 485 | |
| | | *100* | | | *1**0**1* |
| 496 - 20**7** | 289 | | (992 - 406) | 586 | |
| | | *0**1**0* | | | *0**1**0* |
| 496 - 19**7** | 299 | | 992 - 396 | 596 | |
| | | *0**1**0* | | | *0**1**0* |
| 496 - 20**7** | 289 | | (992 - 406) | 586 | |
| | | *100* | | | *1**0**1* |
| 496 - 30**7** | 189 | | (992 - 507) | 485 | |
| 1225 | | *220* | 2222 | | *222* |

(1225 + 220 = **1445**) **/** (2222 + 222 = **2444**) → 999

Last result in Shcherbak's Table: 27 x 37 = 999
(Shcherbak, 1994, Table 1, p. 476)

The relatively small changes in the status of the third perfect number in this Table in relation to Table 2, lead to the large changes in the result. Here it does not follow from the logical square in Figure 2 but follows from two specific arithmetical system-arrangements, presented in Tables 3.1 and 3.2.

**Table 3.1.** *Specific Arithmetical System-Arrangement*
*(SASA I)*

| | | | | | |
|---|---|---|---|---|---|
| 0 | 15 | + | 0 | = | 15 |
| 1 | 1**2**5 | + | 20 | = | 145 |
| **2** | **1225** | + | **220** | = | **1445** |
| 3 | 12225 | + | 2220 | = | 14445 |
| | … | | | | |
| 0 | 15 | + | 9 | = | 24 |
| 1 | 1**4**5 | + | 99 | = | 244 |
| **2** | **1445** | + | **999** | = | **2444** |
| 3 | 14445 | + | 9999 | = | 24444 |
| | … | | | | |

This arithmetical system-arrangement, taken together with the results in Table 3, is a proof more that the key determinant of the Multiplication Table in the decimal number system (3 x 5 = 15) also the determinant of the genetic code (MMR, 2021a, Survey 8 and Tables C2 and C3). The numbers 2 and 4 are emphasized in both subsystems, upper and lower, to indicate the fact that the ratio of number 4 and its half, the number 2, is manual in the genetic code as we have shown in previous works, as also follows from the system-arrangement shown in Table 3.2 (MMR, 2018b, Eq. II in Solution 3, on page 293).

**Table 3.2.** *Specific Arithmetical System-Arrangement*
*(SASA II)*

| | | | | | |
|---|---|---|---|---|---|
| 1 | 2 | + | 22 | = | 24 |
| 2 | 22 | + | 222 | = | 244 |
| **3** | **222** | + | **2222** | = | **2444** |
| 4 | 2222 | + | 22222 | = | 24444 |
| | … | | | | |

For the explanation see Legend of Table 3.1.

**Table 4.** *Arithmetic system corresponding (in the shaded part)
with the mass of 20 amino acid molecules*

| | | | | |
|---|---|---|---|---|
| 2 x (007 x 007) = 98 | 1 x (1070 + 1070) = 2140 | + 2042 | 280 | |
| 2 x (017 x 017) = 578 | 1 x (1170 + 1170) = 2340 | + 1762 | 680 | 400 |
| 2 x (027 x 027) = 1458 | 1 x (1270 + 1270) = 2540 | + 1082 | 1080 | 400 |
| 2 x (037 x 037) = **27<u>38</u>** | 1 x (1370 + 1370) = 27<u>40</u> | + 2 | 1480 | 400 |
| 2 x (047 x 047) = 4418 | 1 x (1470 + 1470) = 2940 | - 1478 | 1880 | 400 |
| 2 x (057 x 057) = 6498 | 1 x (1570 + 1570) = 3140 | - 3358 | 2280 | 400 |
| 2 x (067 x 067) = 8978 | 1 x (1670 + 1670) = 3340 | - 5638 | | |
| . . . | | | | |

The optimal number "2738" corresponds to the mass of 20 amino acid molecules (according to Table 1). Hence the optimality of (prebiotic) selection of 20 protein AAs of standard GC. The difference in relation to the experimental value of only 0.04 is negligible, and is within the limit of experimental error. Nevertheless, the honor goes to the experimenters who determined the atomic masses of the chemical elements that make up amino acid molecules.

# Figures

|     |     |     |     |     |
| --- | --- | --- | --- | --- |
| 189 | 189 | 296 | 485 | 485 |
| 289 | 289 | 297 | 586 | 586 |
| **299** | **299** | 297 | **596** | **596** |
| 289 | 289 | 297 | 586 | 586 |
| 189 | 189 | 296 | 485 | 485 |

**Figure 1.** The mirror symmetry of the quantities contained in Table 1. In the middle area there are the differences of the quantities of molecules mass and the number of nucleons per row, respectively. These differences simultaneously correspond to the quantities of the number of atoms in 61 amino acid molecules within the standard Genetic Code Table (GCT), from the aspect of pyrimidine-purine distinctions: on the left pyrimidine half 32 amino acid molecules have 296 atoms and on the right purine half 29 amino acid molecules have 298 atoms; which means that the mean value of the left and right sides is 297.

| | (0) | (1) | (2) | (3) | (4) | (5) | (6) |
|---|---|---|---|---|---|---|---|
| | 00 | *02* | 04 | 06 | 08 | 10 | 12 |
| | 11 | *13* | 15 | 17 | 19 | 21 | 23 |
| | 22 | *24* | 26 | 28 | 30 | 32 | 34 |
| | 11 | *16* | 21 | 26 | 31 | 36 | 41 |
| | 00 | *05* | 10 | 15 | 20 | 25 | 30 |
| | **44** | **60** | **76** | **92** | **108** | **124** | **140** |
| | | 12 | 14 | 16 | 18 | ***20*** | 22 |
| | | 23 | 25 | 27 | 29 | ***31*** | 33 |
| | | 34 | 36 | 38 | 40 | ***42*** | 44 |
| | | 41 | 46 | 51 | 56 | ***61*** | 66 |
| | | 30 | 35 | 40 | 45 | ***50*** | 55 |
| | | **140** | **156** | **172** | **188** | **204** | **220** |
| | | 22 | 24 | 26 | 28 | *30* | *32* |
| | | 33 | 35 | 37 | 39 | *41* | *43* |
| | | 44 | 46 | 48 | 50 | *52* | *54* |
| | | 66 | 71 | 76 | 81 | *86* | *91* |
| | | 55 | 60 | 65 | 70 | *75* | *80* |
| | | **220** | **236** | **252** | **268** | **284** | **300** |
| | | … | | | | | |

(10) 204, (11) 220 / (15) 284, (16) 300

(220 + 284 = **504**) (204 + 300 = **504**)

**Figure 2.** "The arrangement represents the Table of distinct 2-5 adding (TDA) with starting column 00- 11-22-11-00 which follows from PSN (Periodic system of numbers: Figure A1) in decimal number system by overlapping the real sequence of doubled the first possible triangle in Boolean space (0-1-2) with its mirror image through compression and superposition at the point '22'. In the 10th step we have a realization of the sequence (20-31-42-61-50), the same with the number of atoms in five AAs classes … as it is here presented: all five results in the 10th step are mirror image of the first step." (MMR, 2019, Table A3, p. 30).

|   | 119 |   |   |   |   |
|---|---|---|---|---|---|
| **G 01** | **N 08** | **L 13** | **M 11** | **(33)** |  |
| A 04 | D 07 | K 15 | F 14 | (40) | 120 |
| **V 10** | **S 05** | **R 17** | **Y 15** | **(47)** |  |
| P 08 | T 08 | Q 11 | W 18 | (45) |  |
| **I 13** | **C 05** | **E 10** | **H 11** | **(39)** | 117 |
| G 01 | N 08 | L 13 | M 11 | (33) |  |
| **24**/13 | **18**/23 | **40**/39 | 37/43 | 118/**119** |  |
| (37) | **(41)** | **(79)** | (80) | 117/**120** |  |
|   | 118 |   |   |   |   |

**Figure 3.** <u>The unity of chemism and semiosis</u> (III): "A specific protein amino acids arrangement: The first row is repeated at the bottom, and thus one cyclic system is obtained. There are 117 atoms in two outer columns; at even positions 118, at odd 119; in two inner columns 120 atoms. On the other hand, in the lower half of the Table there are 117 atoms ones more; in the lower diagonally 'wrapped' area 118, and in the upper 119; in the upper half of Table 120 atoms. The repeated four AAs at the bottom of the Table make to achieve a diagonal balance with a difference of only one atom; moreover, to establish a sequence from the series of natural numbers: 117, 118, 119, 120" (MMR, 2017, Table 4, p. 13). [Note: The unity of chemism and semiosis, as found here, is analogous to the such unity in arrangement within Rumer's system of nucleotide doublets. (Cf. footnote 6).]

| | | | | |
|---|---|---|---|---|
| 25 | 38 | 14 | 49 | (**126**) |
| E$_{10}$ 18 N$_{11}$ | R$_{08}$ 22 S$_{13}$ | | | |
| Q$_{09}$ 18 D$_{12}$ | P$_{04}$ 22 F$_{17}$ | | | |
| L$_{06}$ 18 C$_{15}$ | A$_{02}$ 22 W$_{19}$ | | | [1**:**2] |
| | K$_{07}$ 23 T$_{14}$ | | | |
| G$_{01}$ 12 H$_{20}$ | I$_{05}$ 24 M$_{16}$ | | | |
| | V$_{03}$ 25 Y$_{18}$ | | | |
| (**21**) [1**:**3] | 15 | | 48 | (**63**) |

```
      12     18    22   24
          6      4    2
```

(25 – 14 = **11**) (49 – 38 = **11**)
(15 – 14 = **01**) (25 – 15 = **10**)
(48 – 38 = **10**) (49 – 48 = **01**)

**Figure 4.** <u>The unity of chemism and semiosis</u> (IV): "This arrangement of AAs follows from the [logic: first with last and so on]. … The expression of the principle of balancing and nuancing, through the interconnection of the chemical properties of AAs, the number of atoms and the ordinal number of AAs … is self-evident. Confirmation of V. Shcherbak's hypothesis (1994) on the analogy of the amino acid (genetic) code with quantum physics also" (MMR, 2019, Figure 4, p. 22). [*Note 1*: Distances 6–4–2 correspond to the distribution of the number of atoms in 20 AAs as follows. As the number of atoms in the side amino acid chain, the 6 numbers appear once (G-1, A-4, D-7, F-14, R-17, W-18); 4 numbers appear twice (S-5, C-5, V-10, E-10, I-13, L-13, K-15, Y-15); and 2 numbers appear three times (N-8, P-8, T-8, M-11, Q-11, H-11). Hence, follows this pattern: $6_1 – 4_2 – 2_3$ which says the following: six numbers appear once; four numbers appear twice; and two numbers appear three times (MMR, 2018b, Table 4, p. 295). *Note 2*: If we multiply the basic number and the number in the index, we get a new pattern: 6-8-6. If we read this pattern as a three-digit number in the decimal number system (686), then in the case of GC that number can refer to the number of protons, because there are just as many of them in 20 AAs, in their side chains. All together, we can perhaps say (in the future, with a possible positive outcome of the prediction), that not only the principle of self-similarity is at work here, but also the principle of semiosis. For proton number and corresponding data, see in: MMR, 2011, Table 7, p. 830).]

# Surveys

**Survey 1.** *The relationships of quantities contained in Table 2*

$(042 + 300 = \mathbf{342})$ $(220 + 222 = \mathbf{442})$ $[442 - 342 = \mathbf{100}]$

$(220 + 284 = 504)$ $(300 + \boxed{204} = 504)$[a]

$(504 - 342 = \mathbf{1}62)$ $(504 - 442 = \mathbf{0}62)$[b]

$[162 + 062 = 8 \times 28]$[c]  $[162 = 666 - 504]$  $[162 = 2 \times 81]$[d]

$100 : 162 = 0.617 \ldots \approx 0.618 \ldots$ (Golden mean)

---

| | |
|---|---|
| $342 + 442 = \mathbf{784}$ <br> $\mathbf{784} = 28 \times 28$ <br> $\mathbf{384} = 204 + 180$ <br> $(300 + (2 \times 042) = \mathbf{384})$[e] <br> $\mathbf{384} = 496 - (4 \times 28)$ | $(222 = 284 - \mathbf{0}62)$[b] <br> $042 = 204 - \mathbf{1}62$ <br><br> $0.62 \approx 0.618 \ldots$ <br> $1.62 \approx 1.618 \ldots$ |

[a] The four quantities contained in the logical square in Figure 2; two of them are also in Table 2, and other two are indirectly present through a relation with the golden mean. [b] Mediation by golden mean. [c] Mediation with the Golden mean and athe second perfect number, the number 28. [d] The quantities (81 versus 123) as the number of atoms in two classes of AAs, classified according to two classes of the enzymes aminoacyl-tRNA synthetases. [MMR, 2004, Fig. 1, p. 222: "Notice that '81' (as 9 x 9) is the first possible (zeroth) arithmetic square in module 9, and 0-1-2-3 is the first possible (zeroth) logical square (as 00-01-10-11)."] [About the said two classes of AAs one can see in (MMR, 1997a).]  [e] The quantity 384 as the total number of atoms in 20 standard protein amino acids (204 in "bodies" and 180 in "heads") is also found in Plato's Timaeus as the main result of the "harmonization" of geometric progression (MMR, 2011, Table A.2., p. 839).

**Survey 2.** *Golden mean, expressed in different number systems,
in relation to the quantities in Tables 2 and 3*

| Sum of (I) | Golden mean | | |
|---|---|---|---|
| … | | | 504 − 162 = 342 |
| $(770)_8$ | $(1.474 …)_8$ | | |
| | | | 504 − 062 = 442 |
| $(504)_{10}$ | $(1.618 …)_{10}$ | | |
| | | | |
| $(1F8)_{16}$ | $(1.9E3 …)_{16}$ | | 342 = 300 + 042 |
| … | | | |
| "I" → The first friendly pair of numb. | | | 442 = 220 + 222 |
| | | | |
| 222 + 042 = 264 | | 256 = 16 x 16 | |
| 220 + 300 = 264 + **256** | | **256** = 8 x 32 | |
| | | 264 = 8 x 33 | |
| **384** = 256 + ½ 256 | | 297 = 9 x 33 | |
| | | 330 = 10 x 33 | |

For the three equations listed in Remark 3 (very bottom, right), the arithmetical template can be also found as in (MMR, 2018a, Survey 2, positions 8, 9 and 10) or in (MMR, 2021a, Table C10, with corrected cipher errors in line 3 and 27).